\documentclass[twocolumn,showpacs,preprintnumbers,amsmath,amssymb]{revtex4-1}

\usepackage{amssymb}
\usepackage{amsmath}
\usepackage{graphicx}
\usepackage{color}
\usepackage{hyperref}

\newcommand{\be}{\begin{equation}}
\newcommand{\ee}{\end{equation}}
\newcommand{\bea}{\begin{eqnarray}}
\newcommand{\eea}{\end{eqnarray}}
\newcommand{\beal}{\begin{aligned}}
\newcommand{\eeal}{\end{aligned}}

\usepackage{color}

\begin{document}



\title{Entanglement growth during Van der Waals like phase transition}


\author{Hao Xu}
\email{haoxu@mail.nankai.edu.cn}

\affiliation{School of Physics, Nankai University, Tianjin 300071, China}


\date{\today}

\begin{abstract}
We address the problem of describing the coexistence state of two different black holes and Van der Waals like phase transition in Reissner-Nordstr\"om-AdS space-time. We start by a small charged black hole, then introduce a collapsing neutral thin-shell described by Vaidya metric to form a large one. The formation of the large black hole does not change the temperature and free energy of the initial state. We discuss the entanglement growing during the phase transition. The transition is always continuous and the saturation time is determined by the final state. It opens a possibility for studying the holography from excited states to excited states.

\end{abstract}


\maketitle

Black holes are perhaps the most interesting objects in general relativity. They can serve as the perfect tools to test strong gravity. Furthermore, since the groundbreaking work of Bekenstein and Hawking \cite{Bekenstein:1973ur,Bekenstein:1974ax,Hawking:1974sw}, black hole thermodynamics have become a bridge connecting gravity and quantum theory. The phase diagrams of black holes exhibit rich structures and predict phase transitions, such as the Hawking-Page phase transition in Schwarzschild AdS black holes and Van der Waals like phase transition in Reissner-Nordstr\"om-AdS (RN AdS) black holes \cite{Hawking:1982dh,Chamblin:1999tk,Chamblin:1999hg}.

On the other hand, the AdS/CFT correspondence \cite{Maldacena:1997re,Witten:1998qj} has provided us key insight into various systems. The basic idea of AdS/CFT is relating the strongly coupled dual field theory to the classical dynamics of gravity in one higher dimension. The states of the CFT are then dual to bulk geometries with proper boundary conditions. Since the thermal states of the CFT are represented by black holes in AdS, we can use this duality to investigate the black hole phase structures from the boundary theory. If more than one space-time represents an equilibrium state at a given temperature, we can say the CFT has multiple phases. For example, the Hawking-Page phase transition, which describes the phase transition between thermal gas and stable large black holes in Schwarzschild AdS space-time, can be explained as the confinement/deconfinement transition in QCD \cite{Witten:1998zw}.

Van der Waals like phase transition describes a coexistence state of small and large black holes that share the same temperature and free energy, but have different mass and space-time structure. This transition is superficially analogous to the liquid-gas phase transition in Van der Waals fluids. It has been generalized to the extended phase space of black holes, which includes more extensive variables, such as the cosmological constant \cite{Kastor:2009wy,Kubiznak:2012wp}(identified as an effective pressure). Unfortunately, unlike the Hawking-Page transition, the field theory interpretation of Van der Waals like phase transition remains an open question \cite{Johnson:2014yja,Johnson:2013dka,Caceres:2015vsa,Sun:2016til}.

It is natural to ask how to describe the coexistence state of two different black holes and the phase transition process from the gravity side, and even if we can do it, how the field theory on the boundary changes. In this paper we are going to answer these two questions. In the extended phase space in RN AdS space-time, when the pressure $P=-\frac{\Lambda}{8 \pi}$ is lower than critical pressure $P_c$, there is an oscillating part in the isobaric curves on the $T-S$ plane, which should be replaced by an isotherm determined by the Maxwell construction \cite{Lan:2015bia,Xu:2015hba}. It predicts the phase transition temperature and the mass of the small and large black holes. At this temperature, the CFT has multiple phases, so the phase transition can happen only if we can find a way to "transform the space-time". We can start by a small charged black hole, then introduce a collapsing neutral thin-shell described by Vaidya metric propagating from the boundary toward the bulk interior to form the large one. In the outer region there is a large RN AdS black hole, while in the inner region there is a small one. From the field theory point of view, the collapsing thin shell corresponds to the energy rapidly injected to the boundary field. When the equilibrium is reached, the system is described within the grand canonical ensemble. The interesting part is the formation of the large black hole does not change the temperature or free energy of the original small black hole. The initial and final states correspond to different phases of the system. We can say this model describes the coexistence state of two different black holes and a Van der Waals like phase transition.

Earlier studies of holographic dual of black hole formation concentrated on the far from equilibrium process of thermalization in the boundary theory \cite{Balasubramanian:2011ur,Liu:2013qca}. An infalling homogeneous thin mass shell collapses into pure AdS space-time and forms an event horizon, which provided an opportunity to understand the thermalization of nearly inviscid hydrodynamics of relativistic heavy ion collisions at collider energies from AdS/CFT correspondence. During the thermalization process the system is in non-equilibrium background, where thermodynamics quantities may not be well defined \cite{Casas-Vazquez}. Local operators, including energy-momentum tensor and its derivatives, are only sensitive to the metric near the boundary, hence cannot give information on the thermalization process. However, nonlocal operators, such as the entanglement entropy of a region $A$ on the boundary, reach deeper into the space-time and probe into the infrared of the field theory. According to Ryu-Takayanagi formula \cite{Ryu:2006bv,Ryu:2006ef}, the entanglement entropy for a region $A$ of the dual field theory in Einstein gravity can be obtained by calculating the minimal surface $\Sigma$ whose boundary coincides with the $\partial A$: $S_{\textsc{A}}=\frac{Area(\Sigma)}{4G}$, where $G$ is the gravitational constant of the bulk theory.

Notice that our approach is different from the earlier studies ¡°thermalization process¡± in AdS spaces, which describe the transition from pure states to excited states. In this paper, we study the process of transition between different phases with the same temperature.

We start by reviewing the thermodynamics of RN AdS black hole in $d=4$ \cite{Kubiznak:2012wp}. We adapt the unit system by setting the gravitational constant $G$ and speed of light $c$ equal to $1$, then the black hole is represented by the metric
\begin{align}
ds^2=-f(r)\mathrm{d}t^2+\frac{\mathrm{d}r^2}{f(r)}+r^2(\mathrm{d}\theta^2+\sin^2 \theta \mathrm{d}\phi^2),
\label{metric1}
\end{align}
where
\begin{align}
f(r)=1-\frac{2M}{r}+\frac{Q^2}{r^2}+\frac{r^2}{L^2}\,.
\end{align}
The parameter $M$ and $Q$ represent the black hole mass and electric charge respectively, and
$L=\sqrt{-\frac{3}{\Lambda}}$ is the AdS radius.
We can identify the mass
\begin{align}
M=\frac{r_+^3}{2L^2}+\frac{r_+}{2}+\frac{Q^2}{2r_+},
\label{M}
\end{align}
where the horizon $r_+$ is the largest root of $f(r)=0$. The temperature is defined as the surface gravity on the horizon
\begin{align}
T=\frac{f'(r_+)}{4\pi}=\frac{1}{4\pi r_+}\left(1+\frac{3r_+^2}{l^2}-\frac{Q^2}{r_+^2}\right).
\label{T}
\end{align}
The black hole entropy is
\begin{align}
S=\pi r_+^2,
\label{S}
\end{align}
and the electric potential
\begin{align}
\Phi=\frac{Q}{r_+}\,.
\end{align}
Identifying $P=-\frac{\Lambda}{8\pi}=\frac{3}{8\pi L^2}$, the thermodynamical volume is defined as
\begin{align}
V=\bigg(\frac{\partial M}{\partial P}\bigg)_{S,Q}=\frac{4\pi r_+^3}{3}.
\end{align}
Then we can have the first law of black hole thermodynamics
\begin{align}
\mathrm{d} M=T\mathrm{d} S+ V\mathrm{d} P+\Phi\mathrm{d} Q,
\end{align}
and the Smarr formula
\begin{align}
M=2(TS-VP)+\Phi Q.
\end{align}
For a fixed charge $Q$, the free energy can be obtained by calculating the total action \cite{Chamblin:1999hg,Caldarelli:1999xj,Kubiznak:2012wp}. The final result is
\begin{align}
G=\frac{1}{4}\big(r_+-\frac{8\pi}{3}Pr_+^3+\frac{3Q^2}{r_+}\big),
\end{align}
which is also consistent with the Legendre transformation.
The critical(inflection) point is obtained from
\begin{align}
\frac{\partial T}{\partial S}=0\,,\quad \frac{\partial^2 T}{\partial S^2}=0\,,
\end{align}
so we have
\begin{align}
T_c=\frac{\sqrt{6}}{18\pi Q}\,,\quad r_c=\sqrt{6} Q\,,\quad P_c=\frac{1}{96\pi Q^2}\,.
\end{align}
If the pressure is lower than the critical pressure $P_c$, the Maxwell construction is manifested as
\begin{equation}
\begin{aligned}
& T(r_s,P)=T(r_l,P)=T^*\\
& T(r_l,T) \big(S(r_l)-S(r_s)\big)=\int_{r_s}^{r_l}T(r_+,P)\mathrm{d}S(r_+)
\label{law}
\end{aligned}
\end{equation}
where $T^*$ represents the phase transition temperature, and $r_s$, $r_l$ are radius of the small and large black hole respectively. They share the same free energy. Solving the above two equations by relating \eqref{T} and \eqref{S}, we obtain
\begin{equation}
T^*=\sqrt{\frac{8P(3-\sqrt{96\pi{Q^{2}}P)}}{9\pi}},
\end{equation}
and the black hole radius
\begin{equation}
\begin{aligned}
& r_s=\frac{1}{\sqrt{32\pi P}}\bigg(\sqrt{3-\sqrt{96\pi P Q^2}}-\sqrt{3-3\sqrt{96\pi P Q^2}}\bigg)\\
& r_l=\frac{1}{\sqrt{32\pi P}}\bigg(\sqrt{3-\sqrt{96\pi P Q^2}}+\sqrt{3-3\sqrt{96\pi P Q^2}}\bigg)
\label{radius}
\end{aligned}
\end{equation}

\begin{figure}
\begin{center}
\includegraphics[width=0.35\textwidth]{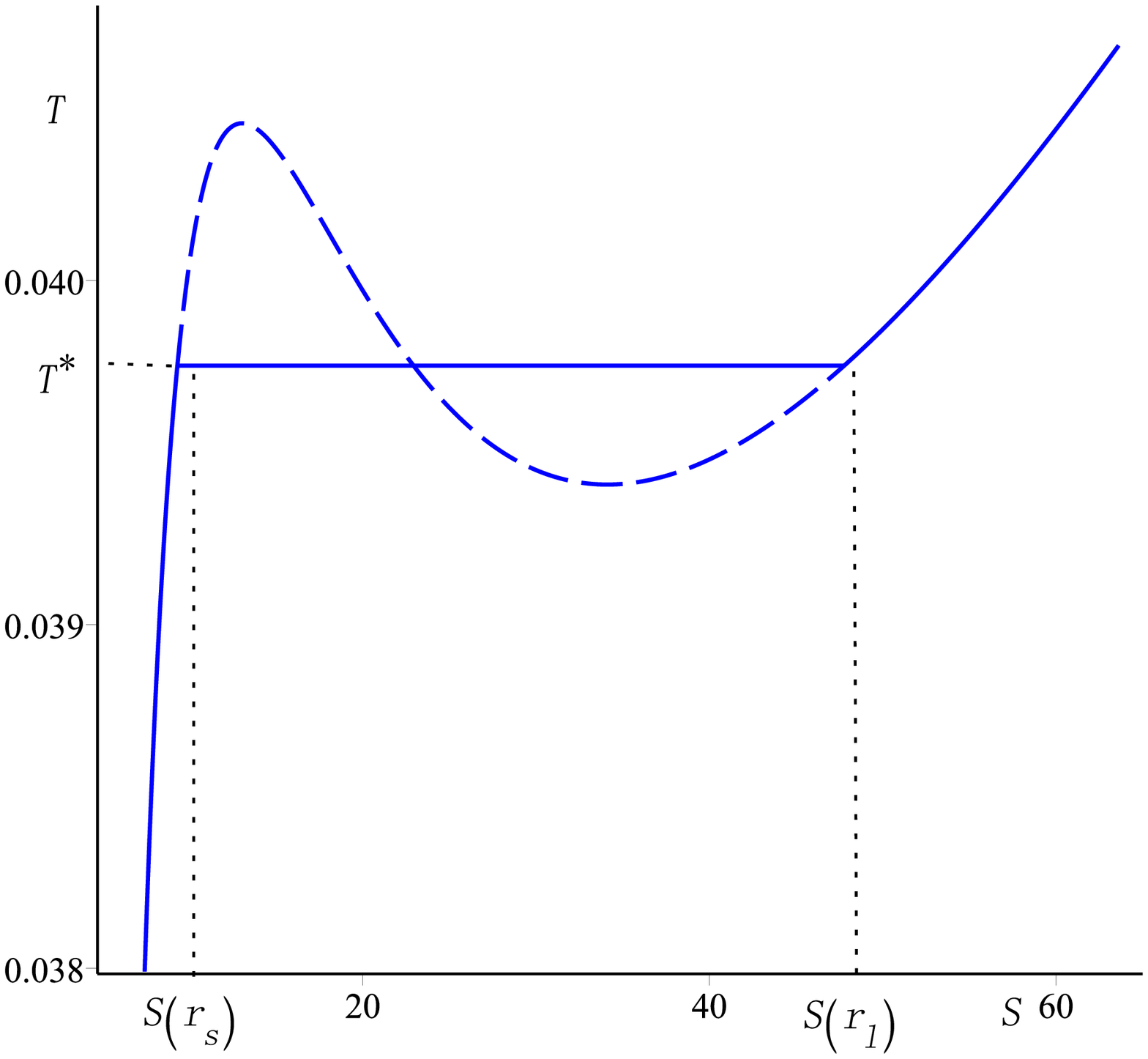}
\includegraphics[width=0.35\textwidth]{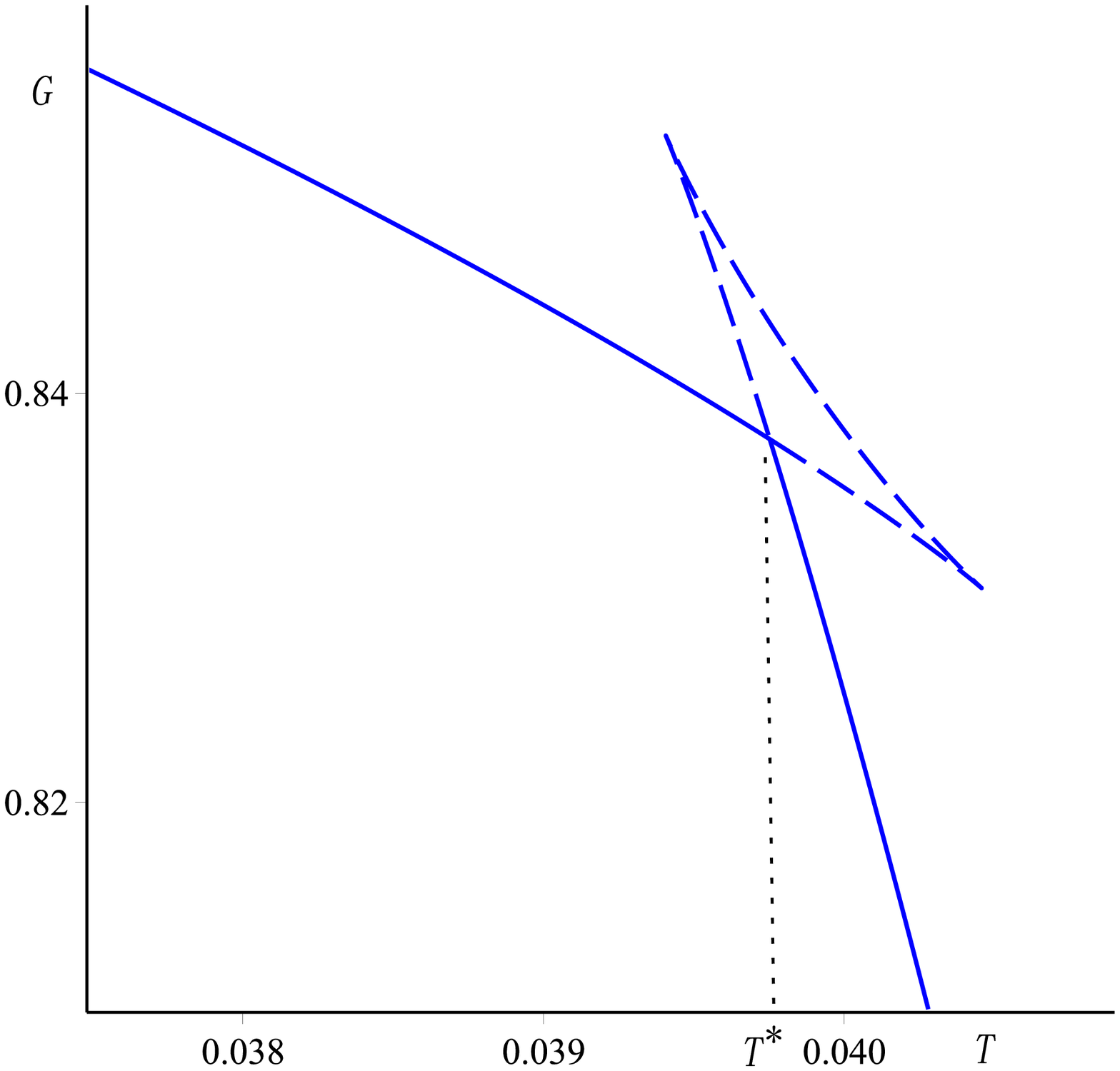}
\caption{Isobaric curves in $T-S$ and $G-T$ planes as $P=0.8P_c$. The dashed part is thermodynamically unstable.}
\label{fig12}
\end{center}
\end{figure}

In Fig.\ref{fig12} we present an example of isobaric curves in $T-S$ and $G-T$ planes as $P=0.8P_c$. Without loss of generality, we set $Q=1$. The dashed part is thermodynamically unstable. The swallow tail in the $G-T$ plane predicts the coexistence state and van der Waals like phase transition.

\begin{figure}
\begin{center}
\includegraphics[width=0.35\textwidth]{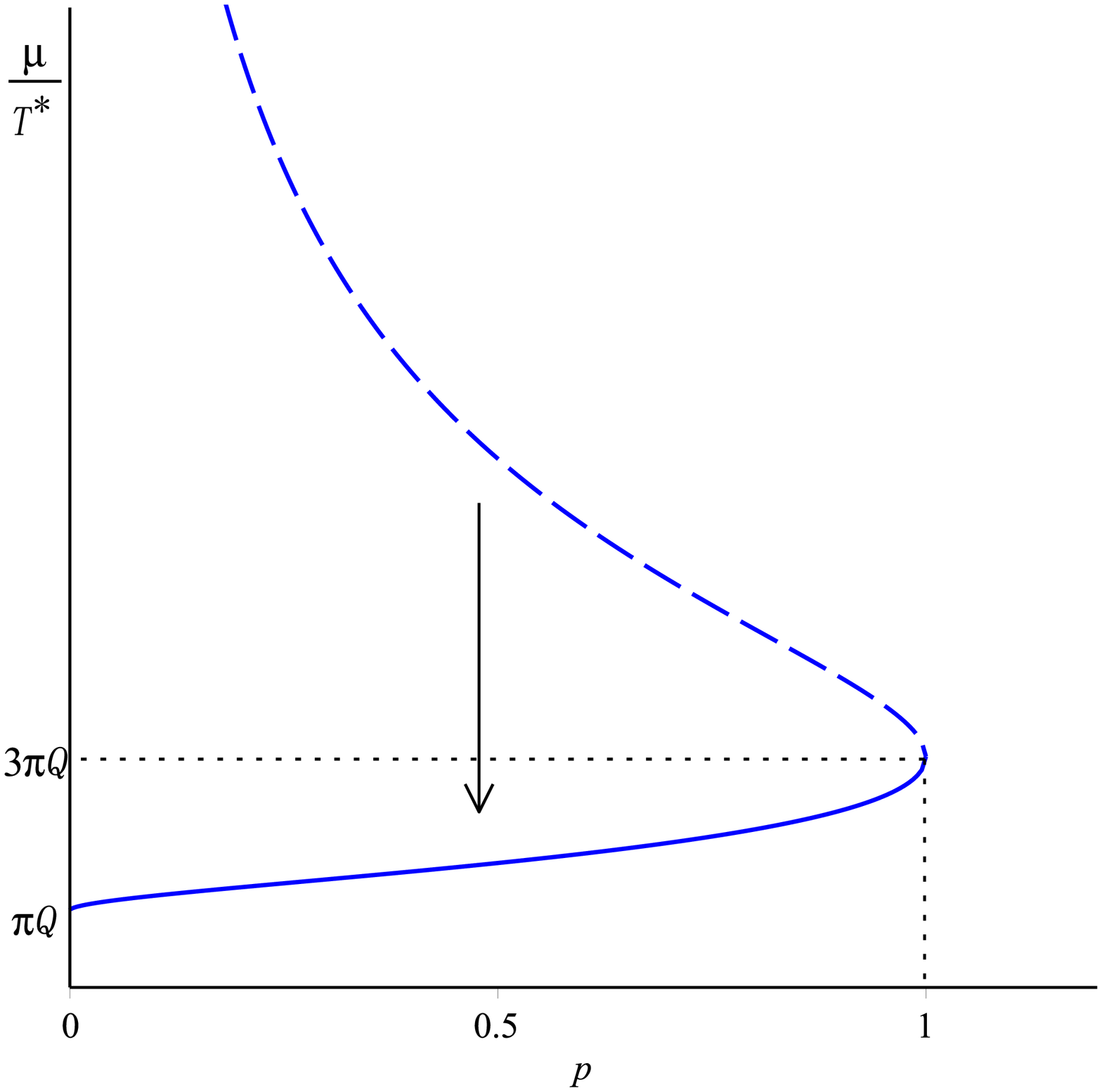}
\includegraphics[width=0.35\textwidth]{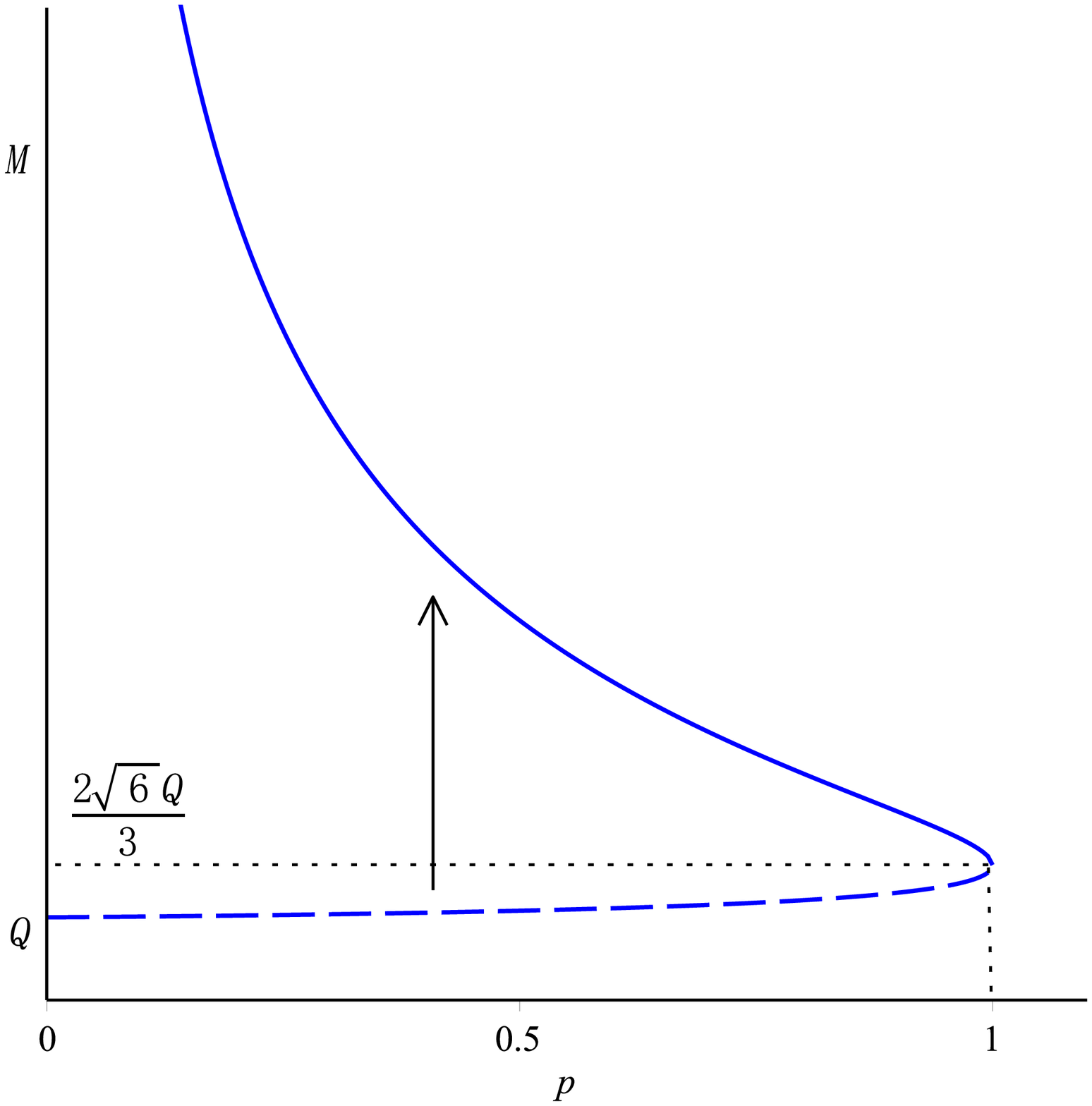}
\caption{$\frac{\mu}{T^*}$ and $M$ as the function of pressure $p=\frac{P}{P_c}$. The dashed and solid parts correspond to small and large black holes during the phase transition.}
\label{fig34}
\end{center}
\end{figure}

The corresponding two black hole mass $M_s$ and $M_l$ can be easily obtained by relating \eqref{radius} and \eqref{M} \cite{Xu:2015hba}. Another interesting quantity is the electromagnetic field, which can be interpreted as the chemical potential $\mu$ in the boundary field theory \cite{Galante:2012pv,Caceres:2012em}. During the phase transition, the chemical potential always decreases. In Fig.\ref{fig34} we present the change of the mass $M$ and chemical potential/ phase transition temperature ratio $\frac{\mu}{T^*}$ during the phase transition. Notice that when pressure goes to zero, the space-time will be asymptotically flat and our analysis no longer applies. In Fig.\ref{fig34} we can observe the mass of the large black hole $M_l$ and the $\frac{\mu}{T^*}$ of the small black hole are both divergent as $P\rightarrow 0$.

Now we focus on the infalling shell geometry described by the Vaidya metric. Introducing the coordinate transformation
\begin{align}
 dv=dt+\frac{dr}{f(r,v)},\quad z=\frac{1}{r},
\end{align}
the RN AdS black hole metric becomes
\begin{align}
ds^2=\frac{1}{z^2}(-f(z,v)z^2dv^2-2dvdz+d \theta^2+\sin^2 \theta d\phi^2).
\label{metric2}
\end{align}
where $f(z,v)$ is represented as
\begin{align}
f(z,v)=1-2M(v) z+Q^2 z^2+\frac{8\pi P}{3 z^2}\,.
\end{align}
The black hole mass $M(v)$ satisfies
\begin{align}
M(v)=M_s, \quad  v<0 \\
M(v)=M_l, \quad  v>0
\label{step}
\end{align}
and we have a zero thickness shell at $v=0$. For the convenience of numerical analysis, we introduce a smooth function
\begin{align}
M(v)=\frac{M_l+M_s}{2}+\frac{M_l-M_s}{2}\tanh\frac{v}{\tilde{v}_0},
\end{align}
where $\tilde{v}_0$ represents a finite shell thickness. When $\tilde{v}_0\rightarrow 0$, we obtain the step function \eqref{step} representing the shock wave.

From the dual field theory point of view, the field theory in our case is on $S^2\times R$, which means it lives in finite volume. However, in the large $N$ limit, there are still infinite number of degrees of freedom in the field \cite{Johnson:2013dka}. The spatial part is on $S^2$.
In coordinates as used in \eqref{metric2}, the subsystem A can be chosen to be bounded by the line of latitude $\theta=\theta_0$, giving a cap shape. Because of the spherical symmetry, the minimal surface bounded by $\theta=\theta_0$ is parameterized as $z(\theta)$ and $v(\theta)$. The induced metric on $\Sigma$ can be written as
\begin{align}
d\tilde{s}^2=\frac{1}{z^2}(-f(z,v)z^2{v'}^2-2v'z'+1)d\theta^2+\frac{1}{z^2}\sin \theta ^2 d\phi^2.
\end{align}
The prime indicates derivative with respect to $\theta$. The holographic entanglement entropy is defined as
\begin{align}
S_A=\frac{1}{4}\int^{\theta_0}_0 d\theta \int^{2\pi}_0  d\phi \frac{\sin \theta}{z^2}\sqrt{1-2z'v'-f(z,v)z^2 v'^2}
\label{HEE}
\end{align}
We can treat the above integrand as the Lagrangian of our system. Since there is an explicit $\sin \theta$
in the Lagrangian, we do not have a conserved quantity associated with the subsystem. However, the equations of
motion can be derived straightforwardly with following boundary conditions
\begin{align}
z(0)=z_*,\quad v(0)=v_*,\quad v'(0)=0,\quad z'(0)=0.
\end{align}
By expanding near $\theta=0$, we have
\begin{align}
z(\varepsilon)=z_*+z_2 \varepsilon^2+O(\varepsilon^3)\\
v(\varepsilon)=v_*+v_2 \varepsilon^2+O(\varepsilon^3)
\end{align}
where
\begin{align}
v_2=\frac{1}{2z_*}
\end{align}
and
\begin{align}
z_2=-\frac{Q^2z_*^3}{2}+M(v)z_*^2-\frac{z_*}{2}-\frac{4\pi P}{3 z_*}
\end{align}
We use the above expansion near $\theta=0$ to solve the equations of motion. For a given $\theta_0$ we can generate the desired minimal surface.
At the boundary we have
\begin{align}
z(\theta_0)=z_0,\quad v(\theta_0)=t.
\end{align}
In principle the AdS boundary should be taken as $z(\theta_0)=0$. However, this will lead to a divergent minimal surface area. Here we take a small number $z_0$ as cutoff.

\begin{figure}
\begin{center}
\includegraphics[width=0.35\textwidth]{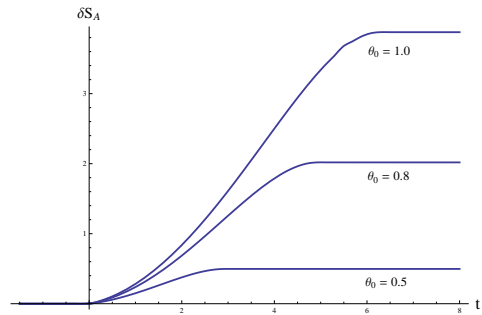}
\caption{Relationship between $\delta S_A$ and $t$ with different $\theta_0$ in $Q=1$, $\tilde{v}_0=0.01$ and $P=0.8P_c$.}
\label{fig5}
\end{center}
\end{figure}

In our case both the initial and final states are excited. Since the value of the minimal surface depends on explicit value of the cutoff $z_0$, we can define the renormalized holographic entanglement entropy by subtracting the initial part from the small black hole
\begin{align}
\delta S_A=S_A(M(v),\theta_0)-S_A(M_s,\theta_0),
\end{align}
so that $\delta S_A$ starts at zero in the infinite past. It is important to study how the $\delta S_A$ evolves as a function of time in different $\theta_0$.  In Fig \ref{fig5} we present the relationship between $\delta S$ and $t$ as we change the $\theta_0$. Here we fix $Q=1$, $\tilde{v}_0=0.01$ and choose $P=0.8P_c$. There are several things to be noted here. Firstly, small systems reach equilibrium faster than the larger ones. This brings no surprise. After the global quench, our system evolves to the final state dual to the large black hole in the bulk. The strongly coupled gauge theory exhibits a topdown mechanism, so the UV modes reach equilibrium first. Secondly, the evolution strongly depends on $\theta_0$. Thirdly, there is no swallow tail behavior near the saturation, which means the transition is always continuous.

We denote $\tau$ as the saturation time. When $t\geq \tau$ the holographic entanglement entropy stops growing and the minimal surface lies entirely in the large black hole region. For a continuous transition, this means at $t=\tau$ the tip of the minimal surface grazes the shell, so the saturation time can be calculated directly as $\tau(\theta_0)=\int^{z_*}_{z_0}\frac{dz}{f(z)z^2}$, where $z_*$ is determined by $\theta_0$. We can obtain that the saturation time only depends on the \emph{final state}. Regardless of the initial state, as long as the transition is continuous, the $\tau$ is uniquely determined by the final state.

\begin{figure}
\begin{center}
\includegraphics[width=0.35\textwidth]{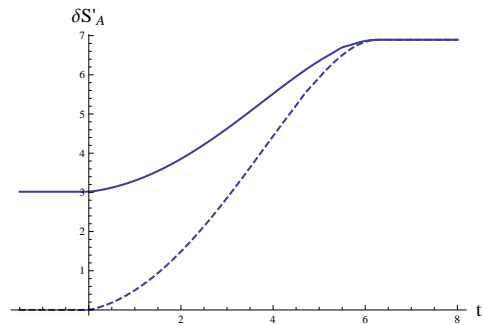}
\caption{Renormalized holographic entanglement entropy $\delta S_A'$ as a function of $t$. The top line describes the phase transition in $Q=1$, $\tilde{v}_0=0.01$, $\theta_0=0.8$ and $P=0.8P_c$. The bottom line describes the thermalization process with the pure AdS space-time as the initial state. They share the same final state and saturation time.}
\label{fig6}
\end{center}
\end{figure}

We introduce a new kind of renormalized holographic entanglement entropy $\delta S_A'$ by subtracting the part from pure AdS space-time. In Fig. \ref{fig6} we compare the above phase transition process with the thermalization process. In the latter one we start in pure AdS space-time, then introduce a charged shell to form an exact same large black hole as in the phase transition process. Then the two kinds of quench have the same final state. We can observe they saturate at the same time, and the renormalized holographic entanglement entropy grows \emph{slower} from excited states than from pure states.

\begin{figure}
\begin{center}
\includegraphics[width=0.35\textwidth]{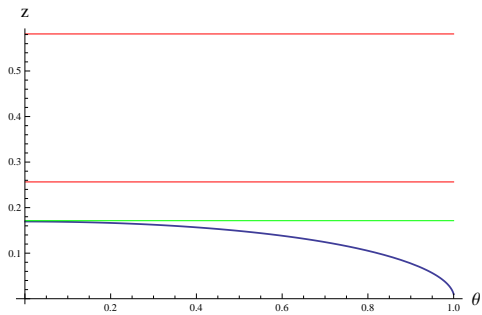}
\caption{Curve $z(\theta)$ at $t=\tau$ in $P=0.8P_c$ and $\theta_0=1.0$.}
\label{fig7}
\end{center}
\end{figure}

In Fig.\ref{fig7} we present an example of curve $z(\theta)$ at $t=\tau\approx6.2781$ in $P=0.8P_c$ and $\theta_0=1.0$. The two red lines $z\approx0.5813$ and $z\approx0.2564$ are the small and large black hole horizon. The green line is the neutral shell. We can observe $z_*$ coincides with the shell at $z\approx0.1693$. Regardless of whether there is a small black hole in the initial state or not, the saturation time $\tau(\theta_0)=\int^{z_*}_{z_0}\frac{dz}{f(z)z^2}$ is only determined by final state.

\begin{figure}
\begin{center}
\includegraphics[width=0.35\textwidth]{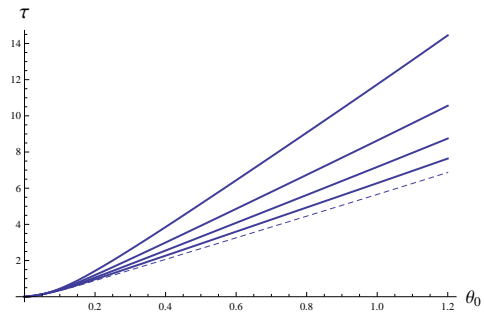}
\caption{Relationship between saturation time $\tau$ and $\theta_0$ for various $P$. From bottom to top $P$ decreases. The top line and bottom dashed line correspond to $P=0.2P_c$ and $P\rightarrow P_c$ respectively. }
\label{fig8}
\end{center}
\end{figure}

We show in Fig.\ref{fig8} the behavior of the saturation time for various $\theta_0$ and $P$. The dashed line corresponds to $P\rightarrow P_c$. At $P=P_c$ the initial and final states coincide, hence there will be no phase transition. From bottom to top the pressure decreases. When $P\rightarrow 0$ the mass and radius of the large black hole are both divergent, so our construction and choice of cutoff will fail. However, since the saturation time $\tau(\theta_0)=\int^{z_*}_{z_0}\frac{dz}{f(z)z^2}$ is divergent only if $z_*\rightarrow z_l$, we can suspect the $\tau$ still remains finite when $P$ is very small. At large $\theta_0$ the slope is almost linear.

\begin{figure}
\begin{center}
\includegraphics[width=0.35\textwidth]{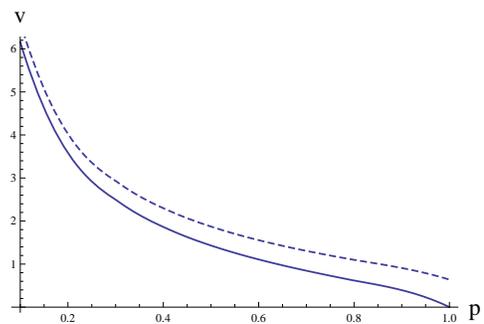}
\caption{Relationship between the average velocity $v$ and pressure $p=\frac{P}{P_c}$ in $\theta_0=1.0$. The dashed line corresponds to the thermalization process with pure AdS space-time as the initial state. The solid line corresponds to the phase transition process. They share the same final state.}
\label{fig9}
\end{center}
\end{figure}

We define the average velocity by calculating the increasing of holographic entanglement entropy over the saturation time $\tau$ during the process. In Fig.\ref{fig9} we present the example of relationship between average velocity $v$ and the pressure $p=\frac{P}{P_c}$ in $\theta_0=1.0$. The dashed and solid lines correspond to the thermalization process and phase transition process which share the same final state. It can be observed the average velocity in the thermalization process provides an upper bound for all the processes with the same final state.

Although in this paper we concern the RN AdS black holes in $d=4$, the exploration can be extended to other dimensions and gravity models, where the phase transition and physics of holographic thermalization can become richer. Other nonlocal operators, such as two-point function and Wilson loop, can be investigated. However, we didn't consider the transition from large black holes to small black holes. In order to do so, we have to extract the energy from the large black hole. Maybe new geometry construction should be introduced. In systems without critical behavior, the current paper can also have potential applications for the study of thermalization process from excited states to excited states. If the system starts in a pure state, the evolution of the system is unitary and the holographic entanglement entropy of $A$ and its complement $B$ are same. When we start in the excited states, $S_A\neq S_B$. In this paper we stay away from strongly thermal regime by choosing $\theta_0$ to be small. However, when $\theta_0$ becomes larger, the numerical analysis we apply will fail and the thermal quench may change the topology of the minimal surface \cite{Azeyanagi:2007bj,Hubeny:2013gta}. This is a non-trivial problem. We hope to be able to address these questions in the near future.

\begin{acknowledgments}

Hao Xu would like to thank Yuan Sun and Liu Zhao for useful discussions. This work is supported by the National Natural Science Foundation of China under the grant No. 11575088.

\end{acknowledgments}

\providecommand{\href}[2]{#2}\begingroup\raggedright\endgroup

\end{document}